\documentclass[aps,prb,twocolumn,superscriptaddress]{revtex4}

\usepackage{graphicx}

\begin{document}


\title{Fano-Kondo interplay in a side-coupled double quantum dot}

\author{S. Sasaki}
\email[]{satoshi@nttbrl.jp}
\affiliation{NTT Basic Research Laboratories, NTT Corporation, Atsugi, Kanagawa 243-0198, Japan}

\author{H.~Tamura}
\affiliation{NTT Basic Research Laboratories, NTT Corporation, Atsugi, Kanagawa 243-0198, Japan}

\author{T. Akazaki}
\affiliation{NTT Basic Research Laboratories, NTT Corporation, Atsugi, Kanagawa 243-0198, Japan}

\author{T. Fujisawa}
\affiliation{Department of Physics, Tokyo Institute of Technology, Meguro, Tokyo 152-8551, Japan}

\date{\today}

\begin{abstract}
We investigate low-temperature transport characteristics of 
a side-coupled double quantum dot where only one of the dots is directly
connected to the leads.
We observe Fano resonances, which arise from interference
between discrete levels in one dot and the Kondo effect,
or cotunneling in general,
in the other dot, playing the role of a continuum.
The Kondo resonance is partially suppressed by destructive Fano interference,
reflecting novel Fano-Kondo competition.
We also present a theoretical calculation based on the tight-binding
model with slave boson mean field
approximation, which qualitatively reproduces the experimental findings.
\end{abstract} 


\maketitle

\newcommand{\tk}{T_{\rm K}}
\newcommand{\vsd}{V_{\rm sd}}

\begin{figure}
\includegraphics{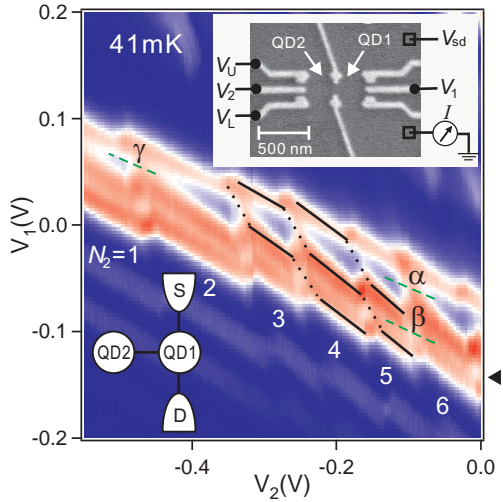}%
\caption{
(Color online)
Color-scale plot of the observed conductance as functions
of the plunger gate voltages, $V_1$ and $V_2$, at 41 mK.
Blue corresponds to zero conductance, white to 20~$\mu$S,
and red to 40~$\mu$S. 
A Kondo valley for QD1 is marked with a triangle.
Solid and dotted lines denote charge state transitions 
on QD1 and QD2, respectively.
The upper inset shows a scanning electron micrograph 
of the DQD device together with a schematic of the measurement setup. 
Current, $I$, is measured across QD1.
The lower inset shows a schematic of side-coupled DQD geometry.
\label{f1}}
\end{figure}

Large tunability of electronic states in semiconductor quantum dot (QD) 
systems has unveiled rich quantum transport phenomena in recent years.
In particular, the Kondo effect \cite{Kondo}
has been extensively investigated as archetypal many-body physics,
and its interplay with Fano interference \cite{Fano} is of great interest
from the viewpoint of tuning spin correlation by quantum coherence.
A QD having an odd number of electrons with spin $S=1/2$
displays the Kondo effect at temperature lower than $\tk$ (the Kondo temperature)
as a result of  spin singlet formation between the localized magnetic moment 
in the QD and the delocalized electrons in the reservoirs. 
Then, the conductance is enhanced at Coulomb blockade regions 
and a zero-bias peak appears
in the differential conductance versus source-drain bias characteristics,
following the formation of a many-body local density of states peak at the Fermi energy
\cite{GGNat,Sara,Schm98,wil00}.

When a second QD (QD2) 
is tunnel-coupled to the side of the first QD (QD1) 
exhibiting the Kondo effect, thus forming a T-shaped geometry 
as schematically shown in the lower inset to Fig.~\ref{f1},
a variety of spin correlation phenomena are expected,
such as the two-stage Kondo effect, the competition with
the inter-dot spin singlet formation,
and the influence by the Fano interference
\cite{Kim,Takazawa,Corn,Tanaka,Wu,Zitko,Chung}.
In contrast to the large number of theoretical studies in the literature,
 very few experiments have been reported on such 
a side-coupled double quantum dot (DQD). 

In this Letter, we report low-temperature transport properties of a side-coupled DQD.
At charge state transitions on QD2, we observe Fano resonances
arising from interference between discrete levels in QD2 
and the cotunneling process in QD1, which serves as a continuum. 
In agreement with the Fano formalism,
the asymmetric resonance line shape is found to evolve systematically with the magnitude of the 
non-resonant channel transmission, {\it i.e.} cotunneling conductance,
which can be changed by the gate voltage or magnetic field.
On the other hand, unlike the standard Fano interference,
the magnitude of the resonance is reduced due to the competition with the Kondo effect.
We also present a tight binding calculation incorporating
slave boson mean field approximation, which is
in qualitative agreement with the experiment.

The DQD device is fabricated 
from a GaAs/Al$_x$Ga$_{1-x}$As heterostructure 
with two-dimensional electron gas (2DEG) located 90 nm from the surface. 
Several Schottky gate electrodes are deposited on the wafer 
to form two parallel QD's as shown in the upper inset to Fig.~\ref{f1}.
Direct connection between QD2 and 2DEG reservoir is lost
by applying sufficiently negative gate biases, $V_{\rm U}$ and $V_{\rm L}$.
Then, the device
operates as a side-coupled DQD with the only current path through QD1.
In this geometry, complications arising from multiple current paths can be avoided,
and even when QD2 has a local magnetic moment, 
it is not screened by the external leads due to the Kondo effect.
Therefore, the influence of QD2 on QD1 transport, or correlated 
transport of the coherent DQD system as a whole, is expected to be
 more straightforwardly investigated than in a serial or parallel DQD.
Transport measurement is performed with a standard lock-in technique 
at temperatures down to $T \sim 40$~mK in a dilution refrigerator. 
The source-drain bias $\vsd = V_{\rm dc} + V_{\rm ac}$, 
where $V_{\rm dc}$ is a DC offset and AC excitation (about 13 Hz) 
$V_{\rm ac} = 5$~$\mu$V. 

Figure \ref{f1} shows a color-scale plot of the linear conductance
as functions of the plunger gate voltages, $V_1$ and $V_2$.
Three strong Coulomb peaks are observed when the electron number 
in QD1, $N_1$, changes by one. 
Inter-dot and dot-lead tunnel barriers are tuned in the strong coupling
regime in excess of $\sim 200 \: \mu$eV for the parameter range of interest.
Inter-dot Coulomb interaction is revealed as jumps in the Coulomb peak 
positions whenever the electron number in QD2, $N_2$, changes by one, 
forming a well-known honeycomb stability diagram \cite{wilDQD}. 
We find such jumps disappear at sufficiently negative $V_2$, 
indicating that $N_2 = 0$ there. 
$N_2$ values determined this way are shown in Fig.~\ref{f1}.
Conductance in the lower Coulomb blockade valley (marked with a triangle) 
 is much larger than in the upper one, 
and a zero-bias peak is observed in the $V_{\rm dc}$ dependence of the 
differential conductance. 
The Kondo effect is responsible for these features
with $\tk \simeq 750$~mK as estimated from the 
half width of the zero-bias peak \cite{wil00}.

\begin{figure}
\includegraphics{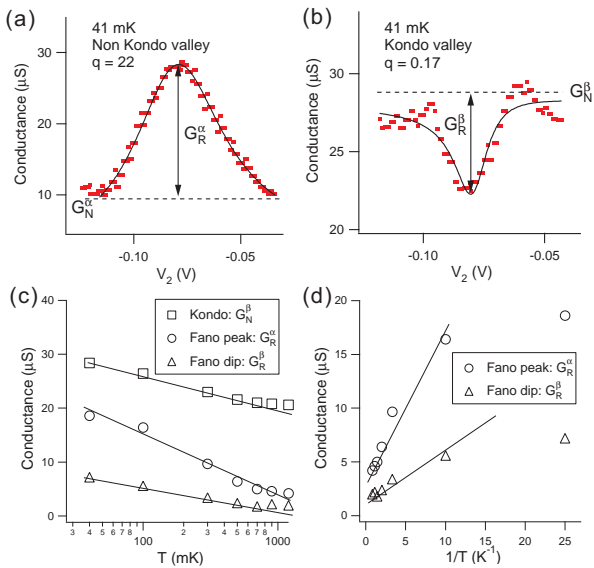}%
\caption{
(a) The observed mid-valley conductance profile across resonance $\alpha$ 
in the non-Kondo valley (dots) fitted with the Fano theory (solid line). 
The dashed line denotes the nonresonant channel conductance, $G_N^{\alpha}$,
and the peak amplitude measured from the dashed line is 
the resonant channel conductance, $G_R^{\alpha}$.
(b) Plot similar to (a) for resonance $\beta$ in the Kondo valley.
(c) Temperature dependence of $G_N^{\beta}$, 
$G_R^{\alpha}$, and $G_R^{\beta}$.
(d) $G_R^{\alpha}$ and $G_R^{\beta}$ plotted as a function of
inverse temperature.
\label{f2}}
\end{figure}

At the boundaries between different $N_2$ ground states where the Coulomb peaks jump,
conductance in the Coulomb blockade region shows 
maxima (minima) at the upper (lower) valley. 
Conductance profiles extracted along dashed lines 
in the middle of the QD1 Coulomb blockade valleys
across QD2 resonances $\alpha$ and $\beta$ (see Fig.~\ref{f1}) 
are plotted in Fig.~\ref{f2}(a) and (b), respectively.
We interpret these conductance modulations in terms of Fano interference
between the indirect transmission via discrete levels in QD2,
and the Kondo effect (cotunneling process) in QD1
playing the role of a continuum.
Data points of conductance $G$ are fitted with the Fano formula (solid line)
\begin{equation}
G (\varepsilon) = G_0 {(\varepsilon + q)^2 \over (\varepsilon^2+1)} + G_1.
\label{fano}
\end{equation}
Here, $\varepsilon \equiv 2(E-E_0)/\Gamma$ is a normalized detuning
with $E$ the energy of an electron, $E_0$ the energy of the resonance,
and $\Gamma$ the level broadening.
$G_0$ is conductance via a nonresonant transmission channel, and
$G_1$ is an offset which is normally ascribed to an incoherent
transmission component.
The dimensionless parameter $q$ is known as Fano's asymmetric parameter 
and characterizes the shape of the resonance line.
Fano interference occurs between transmissions via 
a nonresonant channel (continuum channel) and a resonant one, and their
relative amplitude determines $q$.
When the former dominates, the resonance at $E=E_0$ appears as a dip
with $q \simeq 0$. On the contrary, when the latter dominates, 
the resonance appears as a peak similar to a normal Coulomb blockade
peak with $\left| q \right| \gg 1$.
We define $G_N^i \equiv G_0 + G_1$ ($i = \alpha$, $\beta$ etc.)
as the conductance 
(including the background) ascribed to the nonresonant channel,
{\it i.e.} cotunneling conductance far from the $N_2$ transitions.
We also define resonant channel conductance $G_R^i \equiv (1 + q^2) G_0$. 
When $q = 0$ ($\left| q \right| = \infty$),
$G_R^i$ is the magnitude of the resonance dip (peak) measured with respect to $G_N^i$.
Since cotunneling conductance is relatively small 
($G_N^{\alpha} \simeq 10 \: \mu$S) near the resonance $\alpha$
in the non-Kondo valley, a large resonance {\it peak} is observed
with $q = 22$. On the other hand, a resonance {\it dip} is observed
at $\beta$ in the Kondo valley with $q = 0.17$ and $G_N^{\beta} 
\simeq 29 \: \mu$S \cite{com2}.
A value of $q=1.4$ is obtained with a large asymmetry
in the line shape (not shown) at another resonance $\gamma$ 
in the non-Kondo valley having an intermediate value of 
$G_N^{\gamma} \simeq 25 \: \mu$S. This trend of smaller $q$ for larger $G_N^i$ is
consistent with the Fano formalism because $G_N^i$ reflects 
the transmission amplitude via a nonresonant channel.
Fano resonances have been extensively investigated recently in mesoscopic
systems involving QD's where the nonresonant channel arises from
a quantum wire \cite{Johnson,Koba02,Koba03,Koba04}.
Fano resonances are also observed in a semi-open single QD
without an obvious continuum channel \cite{Gores,Fuhner,Aikawa}.
Although cotunneling has been suggested as a nonresonant channel in Ref.~\cite{Gores},
particular orbital states strongly coupled with the leads 
seem to be more likely sources of nonresonant channels \cite{Aikawa}.
In the present DQD system, cotunneling through QD1 unambiguously serves
as a nonresonant channel, which is corroborated by the device geometry. 

The behavior captured in Fig.~\ref{f2}(b) clearly demonstrates
suppression of the Kondo effect by Fano destructive interference \cite{Wu}.
However, the Kondo zero-bias peak is not completely
suppressed even at the dip region ($V_2 \simeq -0.08$~V).
The amount of this partial suppression of the Kondo peak gives
the dip magnitude, $G_R^{\beta} \simeq 7 \: \mu$S,
while the remaining peak increases the background conductance
to $G_1 \simeq 22 \: \mu$S over a true incoherent background of $\simeq 10 \: \mu$S.
This point is discussed again in the later section dealing with the magnetic field dependence.

Figure~\ref{f2}(c) shows temperature dependence of
$G_N^{\beta}$, $G_R^{\alpha}$, and $G_R^{\beta}$.
$G_N^{\beta}$ is the normal Kondo conductance
showing a linear log$T$ behavior 
as expected. $G_R^{\alpha}$ and $G_R^{\beta}$ seem to change
in a linear log$T$ manner as well, which may suggest their relevance 
to the Kondo physics \cite{Gores}.
The same $G_R^{\alpha}$ and $G_R^{\beta}$ data are plotted
as a function of inverse temperature in Fig.~\ref{f2}(d). 
The high temperature part of the resonant channel conductance decreases 
linearly with temperature,
which suggests thermal broadening of the electron distribution in the leads,
as known for normal Coulomb blockade peaks \cite{Fuhner}.

\begin{figure}
\includegraphics{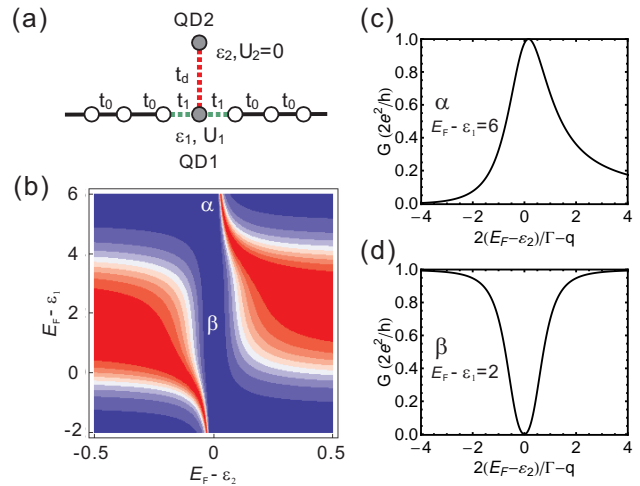}%
\caption{(Color online)
(a) The tight-binding model for the side-coupled DQD.
(b) Color-scale plot of the calculated conductance as functions
of $E_{\rm F} - \varepsilon_1$ and $E_{\rm F} - \varepsilon_2$.
$\alpha$ and $\beta$ correspond to the same symbols in Fig.~\ref{f1}.
(c) Calculated conductance profile at $E_{\rm F} - \varepsilon_1=6$.
(d) Calculated conductance profile at $E_{\rm F} - \varepsilon_1=2$.
Since $\Gamma$ is energy dependent, its value far from 
QD2 resonance is used for normalization.
\label{f4}}
\end{figure}

Next, we present a theoretical calculation of the conductance 
for the side-coupled DQD at $T=0$~K.
We employ the tight-binding model to incorporate relevant parameters
in the present DQD geometry as schematically shown in Fig.~\ref{f4}(a). 
QD1 (QD2) is assumed to have single-particle energy $\varepsilon_1$
($\varepsilon_2$) and the on-site Coulomb energy $U_1$ ($U_2$). 
Inter-dot Coulomb interaction is not taken into account.
An analytical expression for the conductance
equivalent to eq.~(\ref{fano}) is obtained for $U_1=U_2=0$ \cite{tamura09}.
For $U_1 \neq 0$, the Coulomb interaction in QD1 is taken into account by 
the slave-boson mean field approximation. We assume $U_2=0$ to avoid 
calculational complexity,
believing that the essential physics for an individual Fano resonance can be
still captured.  
The conductance is calculated from the transmission probability 
obtained by the S-matrix formalism. 

Figure~\ref{f4}(b) shows the calculated conductance as functions of 
single-particle levels $\varepsilon_1$ and $\varepsilon_2$ with respect to
the Fermi energy, $E_{\rm F}$, for $U_1=4$, $t_1=0.45$, 
and $t_d=0.14$ in units of $t_0$. 
When $\varepsilon_2$ is away from the Fermi level, 
the conductance shows an ordinary Kondo-enhanced conductance
within the QD1 Coulomb blockade valley between $E_{\rm F} - \varepsilon_1 = 0$ and 4. 
As $\varepsilon_2$ approaches $E_{\rm F}$, Fano resonance involving the QD2
discrete level suppresses the Kondo-enhanced conductance in QD1 to zero.
As shown in Fig.~\ref{f4}(d), a symmetric conductance profile with a dip, 
similar to Fig.~\ref{f2}(b), is obtained in the middle of the
QD1 Coulomb blockade valley ($\beta$) at $E_{\rm F} - \varepsilon_1=2$ 
corresponding to $q=0$. On the other hand, an asymmetric peak similar to Fig.~\ref{f2}(a)
is obtained outside the QD1 Coulomb blockade valley ($\alpha$). The conductance profile at 
$E_{\rm F} - \varepsilon_1 = 6$ is shown in Fig.~\ref{f4}(c) corresponding to $q \simeq 5$. 
These results are in qualitative agreement with our experimental results.
The calculated line shape in Fig.~\ref{f4}(d) is slightly different
from the one obtained by putting $q=0$ in eq.~(\ref{fano}) because of the
Kondo correlation. This will be discussed in more detail in a separate publication \cite{tamura09}. 

\begin{figure}
\includegraphics{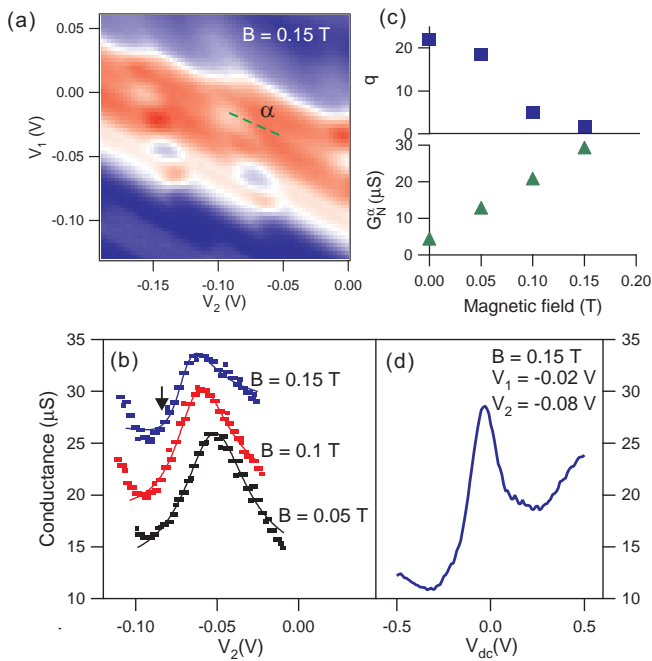}%
\caption{(Color online)
(a) Color-scale plot of the observed conductance as functions
of $V_1$ and $V_2$ at $B=0.15$~T. Color scale is same as in Fig.~\ref{f1}.
(b) Mid-valley conductance profiles across resonance $\alpha$ (dots) 
fitted with the Fano theory (solid line) for various magnetic fields. 
(c) Magnetic field dependence of $q$ and $G_N^{\alpha}$.
(d) A Kondo zero-bias peak at $B=0.15$~T near the resonance minima denoted by an
arrow in (b). The vertical axis is common with (b). 
\label{f3}}
\end{figure}

We finally present experimental results on the magnetic field dependence 
of the Fano resonances. 
When perpendicular magnetic field is applied to a lateral QD,
level crossings occur between different orbital states. Then, the
``Kondo chessboard'' is observed as a result of electron redistribution
within a QD between an inner orbital and an outer one, the latter coupling
more strongly to the leads \cite{stopa03}. 
This means that a non-Kondo valley at zero
field can change to a Kondo valley when its outer orbital holds an odd
number of electrons even though its {\it total} electron number is even.
In fact, the upper non-Kondo valley changes to the Kondo valley at $B=0.15$~T,
as shown in Fig.~\ref{f3}(a). Here, the intensity of the Coulomb valley
conductance is swapped between the upper and lower valleys, and a Kondo
zero bias peak is confirmed in the upper valley, and not in the lower one.
The expected Zeeman energy $\simeq 4 \: \mu$eV at $B=0.15$~T
is much smaller than the half width of the Kondo zero-bias peak $\simeq 85 \: \mu$eV 
(Fig.~\ref{f3}(d)) giving $\tk \simeq 1$~K.
Therefore, suppression of the Kondo effect by magnetic field is negligible.

Mid-valley conductance profiles across resonance $\alpha$ are
plotted in Fig.~\ref{f3}(b) for $B=0.05$~T, 0.1~T, and 0.15~T without an offset.
The application of magnetic field increases $G_N^{\alpha}$, 
as the Kondo effect is activated by the above-mentioned chessboard mechanism.
This is accompanied by the decrease of $q$, {\it i.e.} the 
Fano line shape changes from an almost symmetric peak at zero field
to an asymmetric one at higher field.
These trends, consistent with the Fano formalism,
are summarized in Fig.~\ref{f3}(c), where $q$ and $G_N^{\alpha}$
estimated by fitting the experimental data with eq.~(\ref{fano}), 
are plotted as a function of magnetic field.  
We also note that the conductance minima at the resonance, given by $G_1$ in eq.~(\ref{fano}),
increases with magnetic field, resulting in the reduction of
the resonance amplitude, $G_R^{\alpha}$. 
This might seem strange at first because the increase of conductance due to the Kondo effect,
undoubtedly a coherent transport process, results in the increase of
 $G_1$, which is normally regarded as an incoherent background.
As shown in Fig.~\ref{f3}(d), where $V_{\rm dc}$ dependence of 
differential conductance is plotted for $B=0.15$~T near the resonance minima,
the Kondo zero bias peak is only partially suppressed and the actual incoherent background 
$\simeq 10 \: \mu$S inferred at finite $V_{\rm dc}$
is much smaller than $G_1 \simeq 26 \: \mu$S.
The remaining Kondo peak, which can be regarded as {\it coherent} background,
accounts for the increase of $G_1$ over the true incoherent background
and, in its turn, suppresses the Fano resonance amplitude $G_R^{\alpha}$.
The same situation occurs at $B=0$~T shown in Fig.~\ref{f2}(b).
This mutual suppression demonstrates a novel competition between the Kondo effect 
and the Fano interference.


To summarize, we have investigated transport characteristics 
of the side-coupled DQD both experimentally and theoretically.
Suppression (enhancement) of the conductance is observed in the Kondo 
(non-Kondo) valley of QD1 when charge state transitions occur on QD2.
We demonstrate that these features are Fano resonances where the Kondo effect 
(cotunneling) in QD1 plays the role of a continuum. 
It is found that the Fano destructive interference partially suppresses 
the Kondo resonance, revealing a novel Fano-Kondo interplay.
We have also presented a theoretical calculation based on the tight-binding model
and obtained qualitative agreement with the experimental results.

\begin{acknowledgments}
The authors thank T. Kubo and
Y. Tokura for valuable discussions.
This work was partially supported by SCOPE from
the Ministry of Internal Affairs and Communications of Japan,
and by the Next Generation Super Computing Project,
Nanoscience Program, MEXT, Japan.
\end{acknowledgments}


\begin{thebibliography}{99}
\bibitem{Kondo}
J. Kondo, Prog. Theor. Phys. \textbf{32}, 37 (1964).

\bibitem{Fano}
U. Fano, Phys. Rev. {\bf 124}, 1866 (1961).

\bibitem{GGNat}
D.~Goldhaber-Gordon {\it et al.}, Nature {\bf 391}, 156 (1998).

\bibitem{Sara}
S.~M.~Cronenwett, T. H. Oosterkamp, and L. P. Kouwenhoven, Science \textbf{281}, 540 (1998).

\bibitem{Schm98}
J.~Schmid {\it et al.}, Physica (Amsterdam) {\bf 256B-258B}, 182 (1998).

\bibitem{wil00}
W. G. van der Wiel {\it et al.}, 
Science {\bf 289}, 2105 (2000).

\bibitem{Kim}
T. S. Kim and S. Hershfield, Phys. Rev. B {\bf 63}, 245326 (2001).

\bibitem{Takazawa}
Y. Takazawa, Y. Imai, and N. Kawakami, 
J. Phys. Soc. Jpn. {\bf 71}, 2234 (2002).

\bibitem{Corn}
P. S. Cornaglia and D. R. Grempel, Phys. Rev. B {\bf 71}, 075305 (2005).

\bibitem{Tanaka}
Y. Tanaka and N. Kawakami, Phys. Rev. B {\bf 72}, 085304 (2005).

\bibitem{Wu}
B. H. Wu, J. C. Cao, and K. H. Ahn, Phys. Rev. B {\bf 72}, 165313 (2005).

\bibitem{Zitko}
R. \v{Z}itko and J. Bon\v{c}a, Phys. Rev. B {\bf 73}, 035332 (2006).

\bibitem{Chung}
C. H. Chung, G. Zarand, and P. W\"{o}lfle, 
Phys. Rev. B {\bf 77}, 035120 (2008).


\bibitem{wilDQD}
W. G. van der Wiel {\it et al.}, 
Rev. Mod. Phys. {\bf 75}, 1 (2003).


\bibitem{com2}
Fitting with eq.~(\ref{fano}) is not very good because, 
where the Coulomb peaks jump at $N_2$ transitions,
their tails add to the conductance profile
taken along a straight line not completely parallel to the
Coulomb peaks. 

\bibitem{Johnson}
A. C.~Johnson {\it et al.}, 
Phys. Rev. Lett. {\bf 93}, 106803 (2004).

\bibitem{Koba02}
K.~Kobayashi {\it et al.}, 
Phys. Rev. Lett. {\bf 88}, 256806 (2002).

\bibitem{Koba03}
K.~Kobayashi {\it et al.}, 
Phys. Rev. B {\bf 68}, 235304 (2003).

\bibitem{Koba04}
K.~Kobayashi {\it et al.}, Phys. Rev. B {\bf 70}, 035319 (2004).

\bibitem{Gores}
J. G\"{o}res {\it et al.}, 
Phys. Rev. B {\bf 62}, 2188 (2000).

\bibitem{Fuhner}
C.~F\"{u}hner {\it et al.}, 
cond-mat/0307590.

\bibitem{Aikawa}
H. Aikawa {\it et al.}, J. Phys. Soc. Jpn. {\bf 73}, 3235 (2004).

\bibitem{tamura09}
H. Tamura and S. Sasaki, Physica E, in press.

\bibitem{stopa03}
M. Stopa {\it et al.}, 
Phys. Rev. Lett. {\bf 91}, 046601 (2003).
\end{thebibliography}


\end{document}